\documentstyle[prl,aps,psfig]{revtex}

\begin{document}

\twocolumn

\begin{figure}[t]
\centering
\parbox{0.5\textwidth}{\psfig{figure=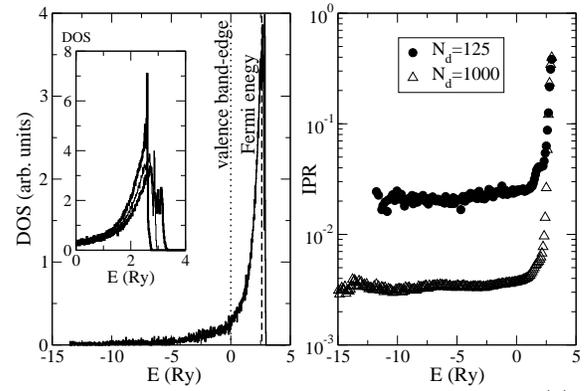,width=75mm,angle=270}}
\caption{\label{fig1} Left: DOS for hopping Hamiltonian with $t(r) <
0$. The inset shows the effect of the AFM coupling ($J=15$ meV). The
$\sigma= \uparrow$ and $\sigma = \downarrow$ bands (thick lines) are
no longer degenerate at low temperatures. For comparison, we plot the
$J=0$ DOS as well (thin line).  Right: dependence of the average IPR
on the energy, In both pictures x=0.0093.  }
\end{figure}

{\bf Berciu and Bhatt reply:} The point raised by the authors of the
Comment\cite{TSO} regarding the hopping term $t(r)$ we use \cite{MB}
to describe 
the impurity band (IB) of holes in Ga$_{1-x}$Mn$_x$As, needs
clarification. A proper description of the hopping  between
impurity states in such alloys is a very complicated and, to our
knowledge, unsettled issue. In our work, we used a hopping parameter
of magnitude corresponding to two isolated s-wave impurities,
with $t(r) < 0$ in the hole representation.  This parametrization
captures two important length scales - the inter Mn-Mn 
distance, and the impurity Bohr radius - present in a
complete model of the experimental system. With our parameters, in the
absence of magnetization, we obtain an impurity band for holes whose
density of states (DOS) is plotted in Fig.\ref{fig1} (left), for a Mn
concentration $x=0.93\%$ using the impurity Rydberg (Ry) as a unit 
(1Ry=112 meV for Mn in GaAs).  The top of the hole impurity band is
3 Ry above the valence band, while the Fermi
energy $E_F$ for holes lies 2.5 Ry above the top of the valence band.
These appear to be reasonable numbers: for a random distribution of
acceptor centers A, the top of the hole band would be expected to
correspond to the ionization potential of the most stable
configuration, which is likely an $A_4$ complex accomodating 4 holes
in the ground impurity state. In a hydrogenic model, this is $\approx
4.5$ Ry\cite{BR}; for centers with strong central cell like Mn in
GaAs, it is likely reduced somewhat.  In the mean-field approach, this
is the situation above $T_c$, while for $T < T_c$, the coupling to the
Mn spins causes the spin up and down bands to split, leaving the
system fully polarized at $T=0$ (see inset).

Additionally, with the parameters used the IB shows a mobility edge
close to $E_F$, as demonstrated by computing the inverse participation
ratio IPR=$\sum_{i}^{} |\phi(i)|^4$, where $\phi(i)$ is the
wavefunction amplitude at site i. For an {\em extended} wave-function,
IPR$\sim 1/N_d$, where $N_d$ is the number of sites of a finite
sample. For {\em localized} wavefunctions, the IPR is independent of
system size. The average IPR for the wavefunctions of our hopping
Hamiltonian are shown in Fig. \ref{fig1} (right) for $N_d=125$ and
1000. The occupied hole states at the top of the hole impurity band
are localized, with relatively high and size-independent IPR, whereas
states of lower energy, which are occupied only at {\em higher} hole
fillings, are extended with IPR's depending on size in the 
expected manner.  Thus, the model captures the proximity of the
Metal-Insulator Transition, seen in experiment. (For $ x = 5\% $, where
the magnetization is much less anomalous, the IB top is again split
a few Ry above the valence band, and is somewhat wider).

On the other hand, as pointed out in the Comment\cite{TSO}, the choice
$t(r) > 0$ in the hole picture, which inverts the IB ($E \rightarrow -
E$), is not suitable because it leads to an unphysical long tail of
the DOS in the gap at the densities of interest, and a consequent low
DOS at $E_F$. Moreover, all the relevant states are extended (there is
no mobility edge).  A more realistic calculation of the IB, including  

{\noindent}the Coulomb potential from the other Mn impurities as well the
charged impurities responsible for the large compensation, and using
the  more complicated structure of a {\em hole} impurity
wavefunction, will likely remove the unphysical tail and yield a DOS
similar to the one used in our calculation, with proximity to a
mobility edge. Already the tail is absent in {\em e.g.} Ref. 3, for a
hydrogenic lattice. Since the nature of compensation in GaMnAs is
still an open question, we opted to use a simple model that gives a
physically acceptable description of the IB.

Our goal in Ref. \cite{MB} was to use this simple IB model to point
out non-trivial effects of disorder in Mn positions on the shape of
the magnetization curve $M(T)$ and the critical temperature $T_{\rm
C}$. To our knowledge, all previously published studies had neglected
this aspect. Studies appearing since support our claim of increased
$T_{\rm C}$ with increased Mn disorder \cite{2}. While we
agree that a better modeling of the impurity band 
and inclusion of the
valence band and other factors such as screening (on a proper, local
scale, taking into account strong charge inhomogeneities)
are necessary to achieve a proper {\em quantitative}
description especially in the metallic regime $x>0.03$, we maintain
that the underlying physics captured by our simple model of
Ref. \cite{MB} is essentially correct.

\end{document}